\documentclass[12pt]{iopart}
\usepackage{graphicx}
\usepackage{iopams}
\usepackage{color}

\begin{document}

\title{Local density of states of the one-dimensional spinless fermion model}
\author{E Jeckelmann}
\address{Institut f\"ur Theoretische Physik, Leibniz Universit\"{a}t Hannover, Appelstra\ss e 2, 
D-30167 Hannover, Germany}
\ead{eric.jeckelmann$@$itp.uni-hannover.de}

\begin{abstract}
We investigate the local density of states of the one-dimensional half-filled
spinless fermion model with nearest-neighbor hopping $t>0$ and interaction $V$
in its Luttinger liquid phase $-2t < V \leq 2t$.
The bulk density of states and the local density of states in open chains
are calculated over the full band width $\sim 4t$ with an energy resolution  $\leq 0.08t$ 
using the dynamical density-matrix renormalization group (DDMRG) method.
We also perform DDMRG simulations with a resolution of $0.01t$
around the Fermi energy to reveal the power-law behaviour 
$D(\epsilon) \sim \vert \epsilon -\epsilon_{\rm F}\vert^{\alpha}$
predicted by the Luttinger liquid theory for bulk and boundary density of states.
The exponents $\alpha$ are determined using a finite-size scaling analysis 
of DDMRG data for lattices with up to 3200 sites.
The results agree with the exact exponents given by the Luttinger liquid theory 
combined with the Bethe Ansatz solution.
The crossover from boundary to bulk density of states is analyzed.
We have found that boundary effects can be seen in the
local density of states at all energies even far away from the chain edges.

\end{abstract}

\pacs{71.10.Fd,71.10.Pm,71.27.+a,73.21.Hb} 

\submitto{\JPCM}

\maketitle

\section{Introduction}

One-dimensional conductors have fascinated physicists for more than 50 
years~\cite{Baeriswyl,Giamarchi}
because they feature unusual properties which set them apart from ordinary 
metals.
Our understanding of ordinary metals is based on the Fermi liquid 
paradigm~\cite{Giuliani}. 
In one dimension, however, this theory fails.
Instead, the low-energy physics of one-dimensional conductors is described 
by the Luttinger liquid paradigm~\cite{Giamarchi,Schoenhammer02,Schoenhammer04}.
The predictions of the Luttinger liquid theory
differ fundamentally from those of the Fermi liquid theory.
For instance, Fermi liquids have a finite density of states 
$D(\epsilon)$ at the Fermi energy $\epsilon_{\rm F}$.
In contrast the bulk density of states of Luttinger liquids
vanishes as a power law at the Fermi energy
\begin{equation}
\label{eq:LLdos}
D(\epsilon) \sim \vert\epsilon-\epsilon_{\rm F} \vert^{\alpha} 
\end{equation} 
where the exponent $\alpha > 0$ depends on the system.
This feature is regarded as one hallmark of a Luttinger liquid.

The density of states $D(\epsilon)$ can be measured experimentally in photoemission
spectroscopy and in scanning tunnelling spectroscopy (STS).
The STS method yields a spatially-resolved 
local density of states (LDOS) with a resolution of a few \AA.
The experimental observation of vanishing densities of states  
has been reported in the photoemission or STS spectrum of various
quasi-one-dimensional conductors such as Bechgaard salts~\cite{Vescoli2000},
the organic charge transfer salt TTF-TCNQ~\cite{Sing2003}, large samples
of single-walled carbon nanotubes~\cite{Ishii2003}, and
the purple bronze Li$_{0.9}$Mo$_6$O$_{17}$~\cite{Hager2005,Wang2006}. 
Consequently, these materials are believed to be
realizations of Luttinger liquids although this 
interpretation remains often controversial.
Only very recently, the power-law behaviour~\eref{eq:LLdos} has been observed 
unambiguously in
the STS and photoemission spectra of gold wires deposited on semiconducting Ge(001) 
surfaces~\cite{Blumenstein2011}.

From a field-theoretical point of view, the low-energy properties
of Luttinger liquids  are very well understood
thanks to powerful methods such as bosonization and
renormalization group~\cite{Giamarchi,Schoenhammer02,Schoenhammer04}.
However, field theory only describes
the low-energy scaling 
$\vert \epsilon - \epsilon_{\rm F} \vert \rightarrow 0$
of various physical quantities such as the density of states.  
Experimentally, this asymptotic behaviour could only be observed
 in a more or less broad window of excitation energies.
Indeed, as real materials are three-dimensional, there is always 
a dimensional crossover~\cite{Baeriswyl,Giamarchi} at low excitation energy
below which one-dimensional physics can no longer be observed.
Moreover, as field-theoretical investigations are based on the linear dispersion of excitations
close to the Fermi energy, there is always a high-energy limit above which
the density of states should deviate from the power-law~\eref{eq:LLdos} because of
the band curvature in real materials~\cite{Imambekov2009}.
Therefore, the theory must be extended to finite-energy scales beyond
the asymptotic behaviour covered by the Luttinger liquid theory 
to facilitate the interpretation of experiments in one-dimensional conductors.

We know that the Luttinger liquid paradigm describes the low-energy
properties of various one-dimensional quantum lattice models for interacting electrons
in their metallic phases,
such as the Hubbard model~\cite{Essler05} away from half filling 
and the spinless fermion model~\cite{Giamarchi,Schoenhammer04}.
Lattice models enable us to study the influence of
finite energy scales such as the curvature and finite width of excitation bands
in one-dimensional conductors.
Some of these models are exactly solvable by the Bethe Ansatz method.
However, it is very difficult to obtain dynamical correlation functions
related to spectroscopic experiments from a Bethe Ansatz solution. 
(The spectral functions
of the spinless fermion models have been investigated only very recently 
using the Bethe Ansatz~\cite{Kohno2010} but the issue of the density of states 
has not been discussed in that work.) 
Therefore, the spectral properties of these models at finite excitation energy
have mostly been determined using numerical methods
for the quantum-many body problem~\cite{Fehske} such as
exact diagonalizations, quantum Monte Carlo simulations and
the density-matrix renormalization group (DMRG) method.
Several power-law divergences ($\sim \vert\epsilon\vert^{\alpha}$ with $\alpha < 0$)
predicted by field theory have been confirmed with these numerical methods.
Yet the power-law singularity~\eref{eq:LLdos} with an exponent $\alpha > 0$
has not been observed univocally
at finite-energy scale in quantum lattice models so far.

The DMRG method is one of the most powerful numerical method
for computing the properties of one-dimensional quantum lattice 
models~\cite{Fehske,Schollwoeck2005}.
The density of states of some lattice models have been investigated using
the original DMRG method~\cite{Schoenhammer2000,Meden2000,Schneider2008}.
In these studies DMRG was used to compute the spectral weight of the lowest 
few eigenstates in order to verify the prediction of the bosonization approach
in the asymptotic low-energy limit.
The dynamical DMRG (DDMRG) is an extension of DMRG which makes possible the
calculation of dynamical correlation functions over their full band width~\cite{Jeckelmann2002,Jeckelmann2008b}.
It yields spectra which are broadened by a Lorentzian of width $\eta$ which sets 
the actual energy resolution.
Over the last decade DDMRG has been used successfully in many studies of spectral properties
in quantum lattice models with energy resolution down to a few hundredths of the bare band 
width~\cite{Jeckelmann2008}. 
Nevertheless, this accuracy has not allowed for a direct observation of 
the power-law behaviour at finite excitation energy until now.
For instance, DDMRG has been used to calculate 
the complete single-particle spectral functions of the Hubbard model
away from half filling with a resolution of $\eta=0.1t$~\cite{Jeckelmann2008}.
The dispersion of holon and spinon branches in the DDMRG spectral functions
agree perfectly with the exact Bethe Ansatz dispersions. 
Yet at the Fermi energy, where the power law~\eref{eq:LLdos} should be seen, 
the momentum-integrated spectral weight barely shows a shallow dip.

In this paper we investigate the density of states of the half-filled
spinless fermion model in its Luttinger liquid phase.
In this model we can study significantly larger systems and thus reach a much better resolution
than in the Hubbard model. Moreover, in the half-filled spinless model any exponent $\alpha \geq 0$ 
can be achieved while in the Hubbard model only $0 \leq  \alpha \leq 1/8$ is possible.   

Both the bulk density of states and the local density of states close to
a chain edge are calculated numerically using the DDMRG method. 
They are determined over the full band width $\sim 4t$ with a 
resolution of $\eta=0.04t$ or $0.08t$. 
Using lattices with up to 1600 sites 
we have also performed DDMRG simulations with a resolution
of $\eta=0.01t$ around the Fermi energy to reveal
the power-law behaviour~\eref{eq:LLdos}
predicted by the Luttinger liquid theory for bulk and boundary density of states.
The exponents $\alpha$ are determined using a finite-size scaling analysis
of DDMRG data for lattices with up to 3200 sites and compared to the predictions of the
Luttinger liquid theory combined with the exact Bethe Ansatz solution.
Finally, we discuss the crossover from boundary to bulk density of states
as one moves away from the chain edges.

Our paper is organized as follows: 
In the next section we introduce the model and method used in this work. 
In the third section we discuss our results for the bulk
density of states while the LDOS close
to a chain edge is analyzed in the fourth section.
Finally, our findings are summarized in the last section.

\section{Models and method}

The one-dimensional spinless fermion model is one of the simplest
realizations of a one-component Luttinger liquid in a lattice model.
It can be interpreted as a system of spin-polarized electrons.
The model is defined by the Hamiltonian
\begin{equation}
\label{eq:hamiltonian}
H = - t \sum_{j=1}^{N-1} \left (c_{j}^{\dag}c^{\phantom{\dag}}_{j+1} +
c_{j+1}^{\dagger}c_{j}^{\phantom{\dag}} \right )
+ V \sum_{j=1}^{N-1}\left (n_{j}-\frac{1}{2} \right ) \left (n_{j+1} - \frac{1}{2} \right ) 
\end{equation}
where $c_{j}^{\dag}$ and $c^{\phantom{\dag}}_{j}$ are the creation and annihilation operators
for a spinless fermion on site $j$, and the density operator is 
$n_{j} = c_{j}^{\dag}c_{j}^{\phantom{\dag}}$. The model parameters are 
the hopping amplitude $t > 0$ between nearest-neighbor sites
and the Coulomb interaction $V$ between particles on nearest-neighbor sites.
The half-filled system corresponds to $N/2$ fermions on the $N$-site lattice. 

The one-dimensional spinless fermion model is exactly solvable by the Bethe 
Ansatz method~\cite{Giamarchi,Schoenhammer04}.
For $-2t<V\leq2t$ its excitation spectrum is gapless and its low-energy 
properties are described by the Luttinger liquid theory.
At half filling the dispersion of elementary excitations~\cite{Caux2011} is given by
\begin{equation}
\epsilon(k) = 2t^{*} \vert \sin(ka)\vert
\label{eq:dispersion}
\end{equation}
with the renormalized ``hopping term''
\begin{equation}
\label{eq:hopping}
t^{*} = \frac{\pi t}{2} \frac{\sqrt{1-\left (\frac{V}{2t} \right )^2}}{\arccos 
\left (\frac{V}{2t} \right )}  .
\end{equation}
The universal properties of a one-component Luttinger liquid are determined 
by two parameters: The velocity of elementary excitations (renormalized Fermi velocity) $v$
and a dimensionless Luttinger parameter $K$.
From the Bethe Ansatz solution we know the relation between these Luttinger liquid parameters 
and the lattice model parameters at half filling,
\begin{equation}
\label{eq:velocity}
v = \frac{2a t^{*}}{\hbar}
\end{equation}
with $t^{*}$ given by~\eref{eq:hopping}  and
\begin{equation}
\label{eq:Kparameter}
K = \frac{\pi}{2} \frac{1}{\pi - \arccos \left ( \frac{V}{2t} \right ) }
\end{equation}
where $a$ is the lattice constant. In the non-interacting chain $V=0$ 
this yields $K=1$ and the usual Fermi velocity $v=v_{\rm F} = 2ta/\hbar$
of the one-dimensional tight-binding model.

The local density of states (LDOS) is defined by
\begin{equation}
D(j,\epsilon) = \cases{
\sum_n \left \vert \left \langle n \left \vert c^{\dag}_j \right \vert 0 \right \rangle \right \vert^2
\delta(\epsilon-E_n+E_0) & for $\epsilon > \epsilon_{\rm F}$  \\ 
\sum_n \left \vert \left \langle n \left \vert c^{\phantom{\dag}}_j \right \vert 0 \right \rangle \right \vert^2
\delta(\epsilon+E_n-E_0)  & for $\epsilon < \epsilon_{\rm F}$}
\label{eq:dos}
\end{equation}
where $\vert n \rangle$ denotes the eigenstates of the Hamiltonian $H$ and 
$E_n$ their eigenenergies in the Fock space. 
The ground state for the chosen number of particles correspond to $n=0$.
The total spectral weight is 
\begin{equation}
\int_{-\infty}^{+\infty} d \epsilon \ D(j,\epsilon) = 1 .
\label{eq:sum}
\end{equation}
As the Hamiltonian~\eref{eq:hamiltonian} is invariant under the particle-hole transformation
$c^{\dag}_j \leftrightarrow  c^{\phantom{\dag}}_j  (-1)^j$, its density of states is symmetric at 
half filling: $D(j,\epsilon) = D(j,-\epsilon)$  with $\epsilon_{\rm F} = 0$.

We use the DDMRG method~\cite{Jeckelmann2002,Jeckelmann2008b}
to compute the density of states~\eref{eq:dos}. 
DDMRG simulations yield the convolution of 
a Lorentzian of width $\eta > 0$ with
the local density of states on a $N$-site lattice
\begin{equation}
D_{\rm DMRG}(j,\epsilon) = \frac{1}{\pi}  \int_{-\infty}^{+\infty} d \epsilon'  
\frac{\eta}{( \epsilon-\epsilon')^2 + \eta^2} \ D(j,\epsilon') .
\label{eq:convolution}
\end{equation}
As we are interested in the density of states in the
thermodynamic limit, we should, in principle, 
extrapolate our numerical data to an infinite lattice size
$N \rightarrow \infty$ and then to a vanishing broadening $\eta \rightarrow 0$.
However, the spectrum of a $N$-site chain is indistinguishable
from the same spectrum in the thermodynamic limit if they
are compared on an energy scale (resolution) $\Delta \epsilon > W/N$.
The parameter $W$ has to be determined empirically
but it is typically of the order of the effective band width
of the spectrum considered (here $W \sim 4t^{*}$).
Thus if we choose a broadening $\eta > W/N$,
DDMRG spectra can be regarded as 
the spectra of infinite systems with the same broadening 
$\eta$, which sets the actual resolution.
In this work we have systematically used 
\begin{equation}
\eta = \frac{16t}{N} > \frac{4t^{*}}{N} .
\label{eq:eta}
\end{equation}

The DMRG algorithm truncates the Hilbert space in an optimal way 
to represent some chosen target states such as the ground state.
Thus DMRG data are affected by truncation errors which depend on the number of
density-matrix eigenstates kept in the calculation~\cite{Fehske,Schollwoeck2005}.
In the DDMRG method the truncated Hilbert space is optimized to represent
the ground state and eigenstates with a given excitation energy $\epsilon$
or lying in a given excitation energy range $[\epsilon_1,\epsilon_2]$
with $\epsilon_2 - \epsilon_1 \lesssim \eta$. The calculations
are carried out independently for each excitation energy. 
We have kept up to 400 density-matrix eigenstates in our simulations. 
Numerical errors are always very small in absolute values
or when they are compared to the total spectral weight~\eref{eq:sum}.
Thus the DDMRG results presented here can be regarded as numerically exact.
This is illustrated in the following sections by comparison with exact 
spectra for non-interacting fermions.
However, as the spectral weight approaches zero at the Fermi energy,
relative errors become unavoidably larger and can be seen as a scattering
of DDMRG data close to the Fermi energy.
In our numerical calculations we use the energy scale $t=1$, the time scale $\hbar/t=1$
and the length scale $a=1$.

\section{Bulk density of states}

\begin{figure}[t]
\centering
\includegraphics[width=8cm]{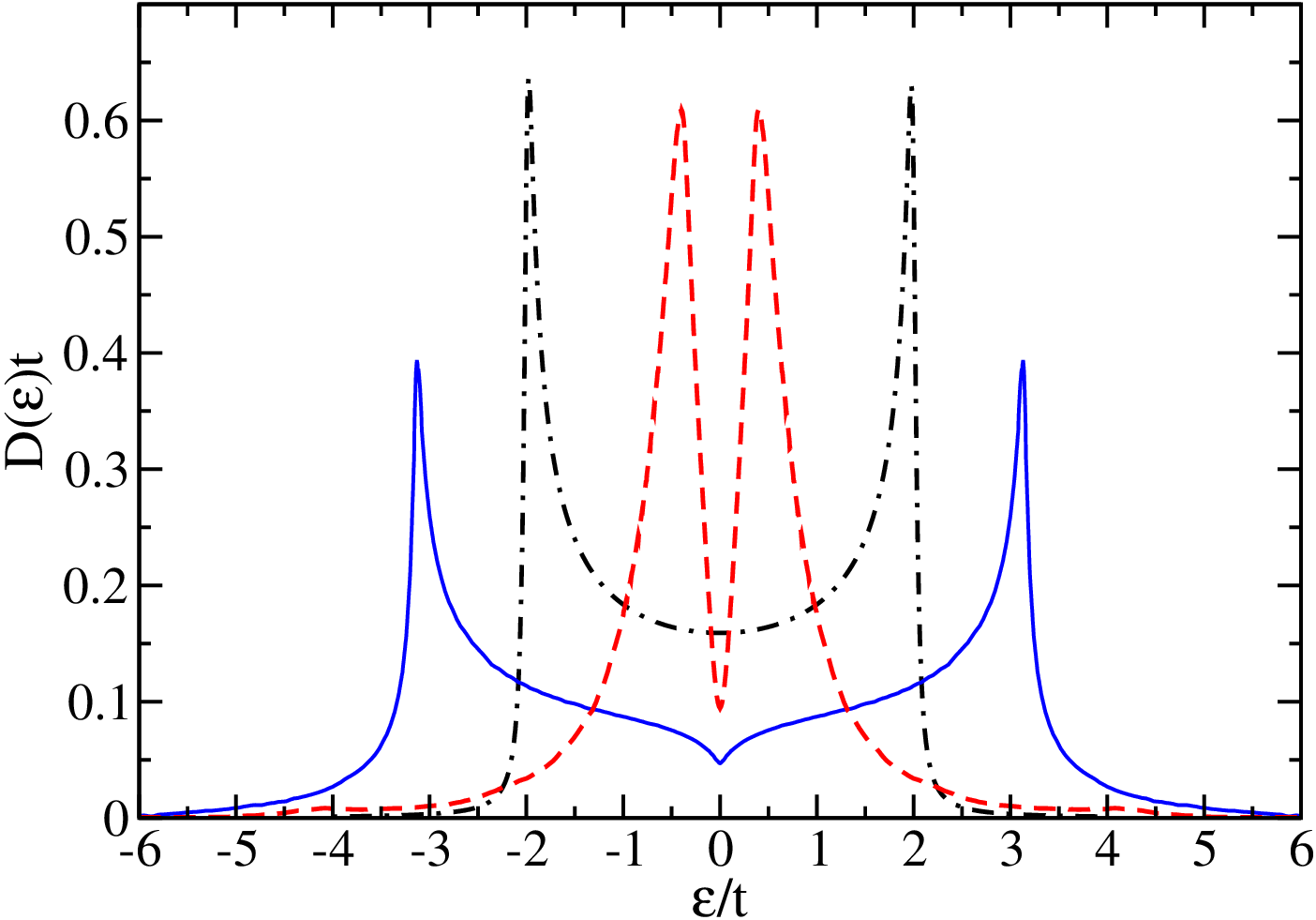}
\caption{\label{fig:bulkdos} Bulk density of states for $V=-1.9t$ (\dashed), $V=0$
(\chain), and $V=2t$ (\full)
calculated in the middle of a 400-site chain ($\eta=0.04t$).}
\end{figure}

In principle, the bulk density of states $D(\epsilon)$ should be calculated as the average 
of the 
LDOS~\eref{eq:dos} over the full lattice in the thermodynamic limit. 
Here we identify the bulk density of states
with the LDOS in the middle of a $N$-site chain, $D(j=N/2,\epsilon)$.  
While this seems intuitively correct for insulators, 
in a Luttinger liquid boundary effects could a priori 
be felt as far as the chain center. 
First, we have verified the validity of this approach
for several exactly solvable non-interacting models.
For instance, the bulk density of states in a non-interacting chain ($V=0$) can be calculated
analytically,
\begin{equation}
D(\epsilon) = \frac{1}{\pi} \frac{1}{\sqrt{4t^2-\epsilon^2}}  
\label{eq:bulkdos0}
\end{equation}
for $\vert \epsilon \vert < 2t$ and $D(\epsilon) = 0$ for
$\vert \epsilon \vert > 2t$.
\Fref{fig:bulkdos} shows the LDOS calculated with DDMRG in the middle of a 400-site 
non-interacting chain
with $\eta=0.04t$. On the scale of this figure 
the DDMRG spectrum is indistinguishable from the exact result~\eref{eq:bulkdos0}
convolved with a Lorentzian of the same width $\eta$.
This demonstrates the accuracy of the DDMRG method and also 
confirms that the scaling of the broadening~\eref{eq:eta} is appropriate for
the density of states of the spinless fermion model.

Additionally, we have checked in interacting systems with DDMRG that 
$D(j,\epsilon)$ depends negligibly on the site position
$j$ close to the chain center and that $D(j=N/2,\epsilon)$ is indistinguishable 
from the momentum-integrated spectral functions~\cite{Jeckelmann2008b}.
Finally, we will examine the crossover from boundary LDOS to bulk LDOS
in more detail in the next section.  
All our results confirm that the DDMRG spectra calculated in the middle of a 
spinless fermion chain with a broadening~\eref{eq:eta}
can be regarded as its bulk density of states in the Luttinger liquid phase.

\begin{figure}[t]
\centering
\includegraphics[width=8cm]{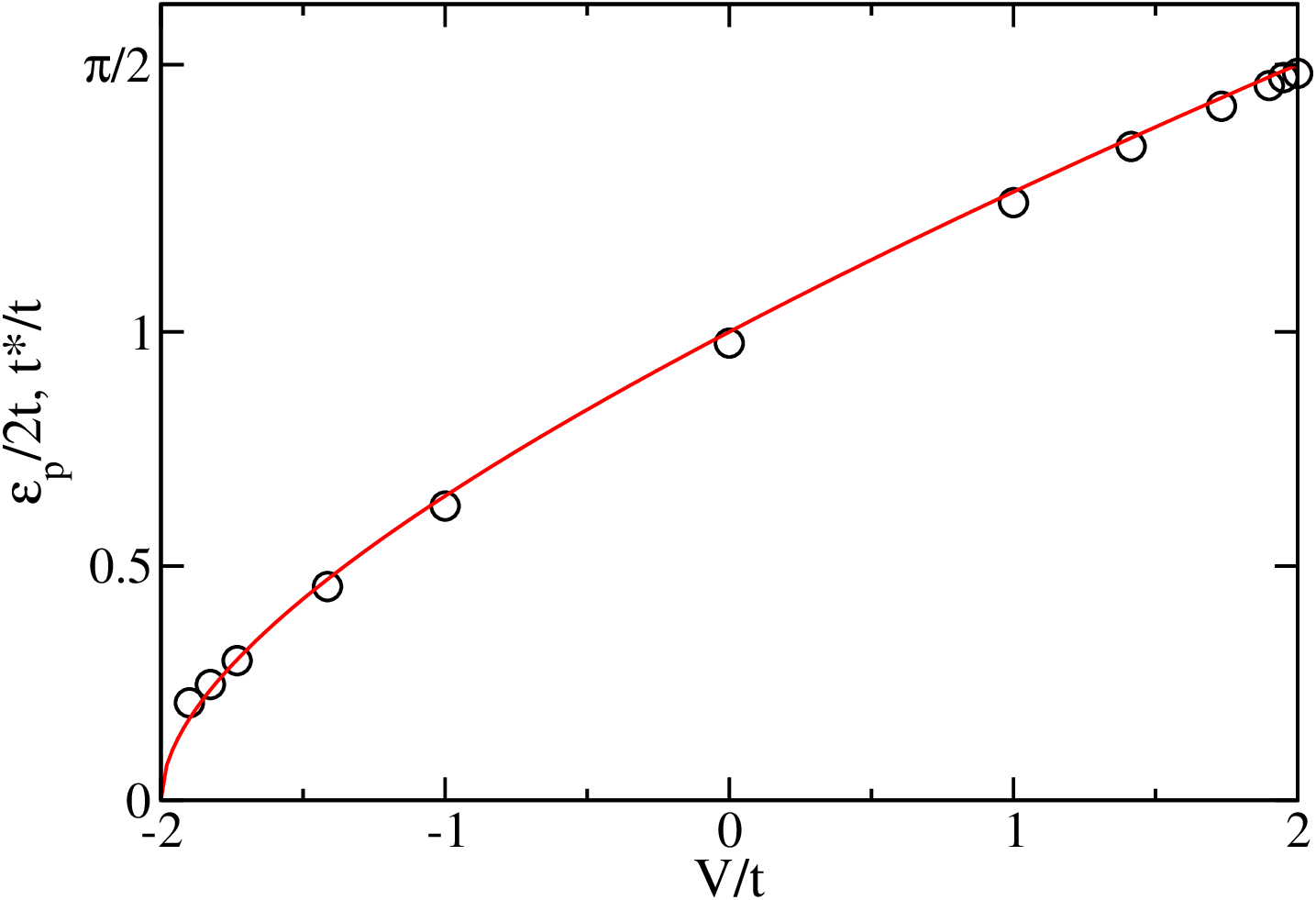}
\caption{\label{fig:bandwidth} Position $\epsilon_{\rm p}/2t$ of the peaks in the bulk 
density of states
calculated with DDMRG in 200-site chains (\opencircle) and 
renormalized  ``hopping term'' $t^{*}/t$ from~\eref{eq:hopping} (\full)
as a function of the interaction parameter $V$.}
\end{figure}

\Fref{fig:bulkdos} shows the bulk density of states over the full band width 
for three different values of the interaction $V$. 
The band edge singularities of the non-interacting 
density of states~\eref{eq:bulkdos0} are seen as two peaks at $\epsilon = \pm 2t$
because of the broadening $\eta$.
Similar peaks are visible at positions $\epsilon = \pm \epsilon_{\rm p}$ for $V \neq 0$. 
The energy $\epsilon_{\rm p}$ increases monotonically with $V$
starting from 0 for $V =-2t$.
Actually, \fref{fig:bandwidth} shows that 
$\epsilon_{\rm p}/2t$ varies exactly as the effective ``hopping term''
$t^{*}$~\eref{eq:hopping} as a function of $V$.
Thus $\epsilon_{\rm p} = 2t^{*}$ and the peak positions correspond to
the edges of the elementary excitation bands 
[i.e., the maximum of the dispersion~\eref{eq:dispersion}]
for all values of $V$.

In \fref{fig:bulkdos} we see some spectral weight beyond $\epsilon_{\rm p}$ 
which is clearly due to the broadening $\eta$ for $V=0$ but 
cannot be explained by this broadening for $V=-1.9t$ and $V=2t$. 
Therefore, excitations which are made up from more than one elementary excitations
contribute to the density of states~\eref{eq:dos} 
at energy $\vert \epsilon\vert > \epsilon_{\rm p}$.

According to the Luttinger liquid paradigm
the bulk DOS should vanish as the power law~\eref{eq:LLdos}
at the Fermi energy. 
The Luttinger liquid theory also
predicts that the exponent is 
\begin{equation}
\alpha= \frac{(K-1)^2}{2K} 
\label{eq:exponent1}
\end{equation}
for a one-component Luttinger liquid.
Thus for the spinless fermion model this exponent can 
be calculated exactly as a function of the interaction parameter $V$
using the Bethe Ansatz relation for the Luttinger parameter~\eref{eq:Kparameter}.
In the non-interacting chain $(V=0)$ this yields $\alpha=0$, which implies
that the bulk DOS does not vanish but approaches a constant value at the Fermi energy,
while one gets $\alpha > 0$ for interacting chains $(V\neq 0)$.

\begin{figure}[t]
\centering
\includegraphics[width=8cm]{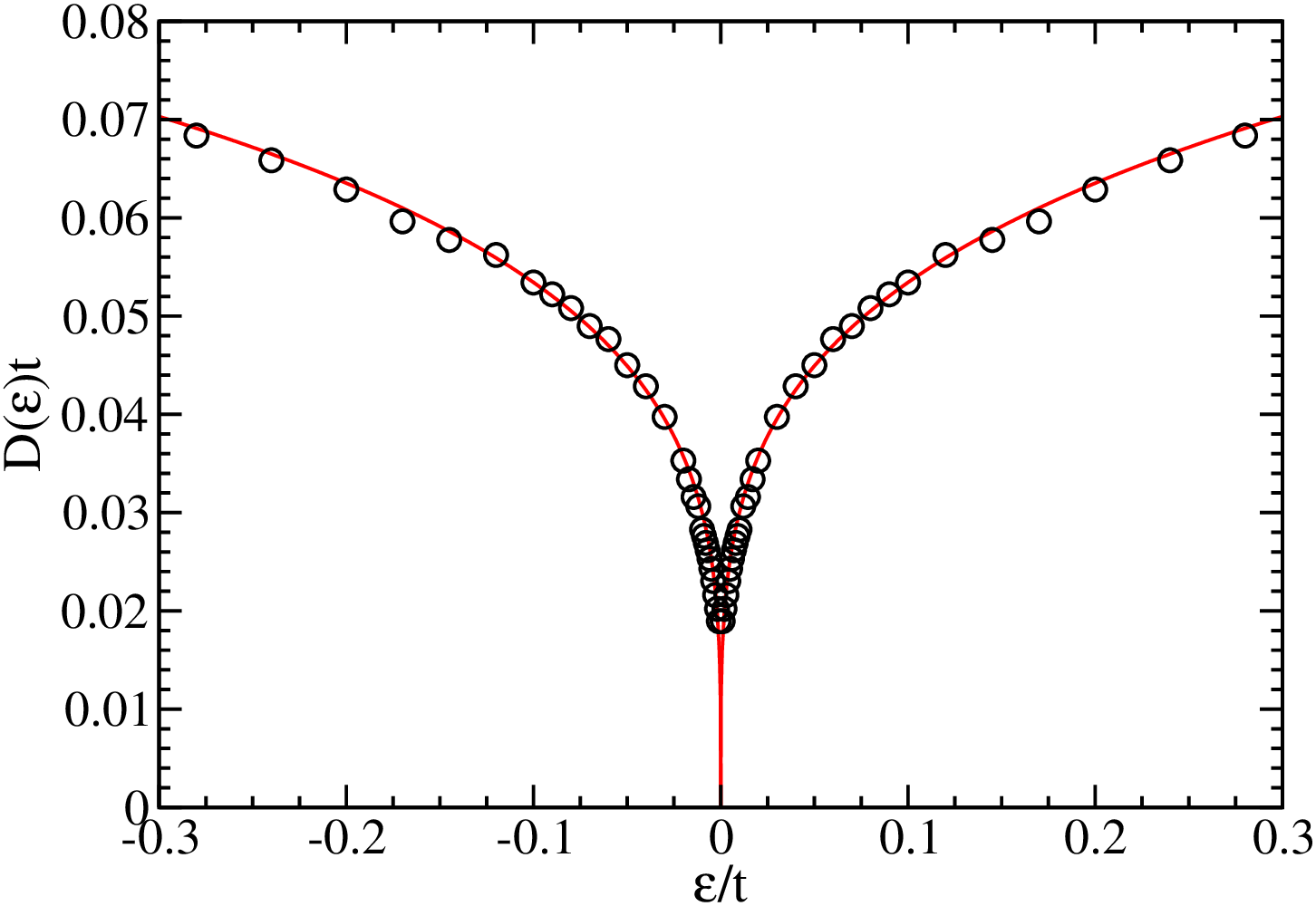}
\caption{\label{fig:expbulkdos} Bulk density of states around the Fermi energy for
$V=2t$ calculated with DDMRG in a 1600-site chain ($\eta=0.01t$).
The solid line is the power law~\eref{eq:LLdos} with the exact exponent 
$\alpha=\frac{1}{4}$.}
\end{figure}

The power-law behaviour is not visible on the scale of \fref{fig:bulkdos}.
The vanishing of the density of states is just suggested by narrow minima 
which are seen at $\epsilon=0$ for interacting fermion systems. 
\Fref{fig:expbulkdos} shows the bulk density of states around
the Fermi energy at a higher resolution ($\eta=0.01t$) for $V=2t$. 
The power law~\eref{eq:LLdos} is also shown
with the exact exponent $\alpha=\frac{1}{4}$ for $V=2t$.
Clearly, the DDMRG data agree with the power law over the energy range shown in this figure.

\begin{figure}[b]
\centering
\includegraphics[width=8cm]{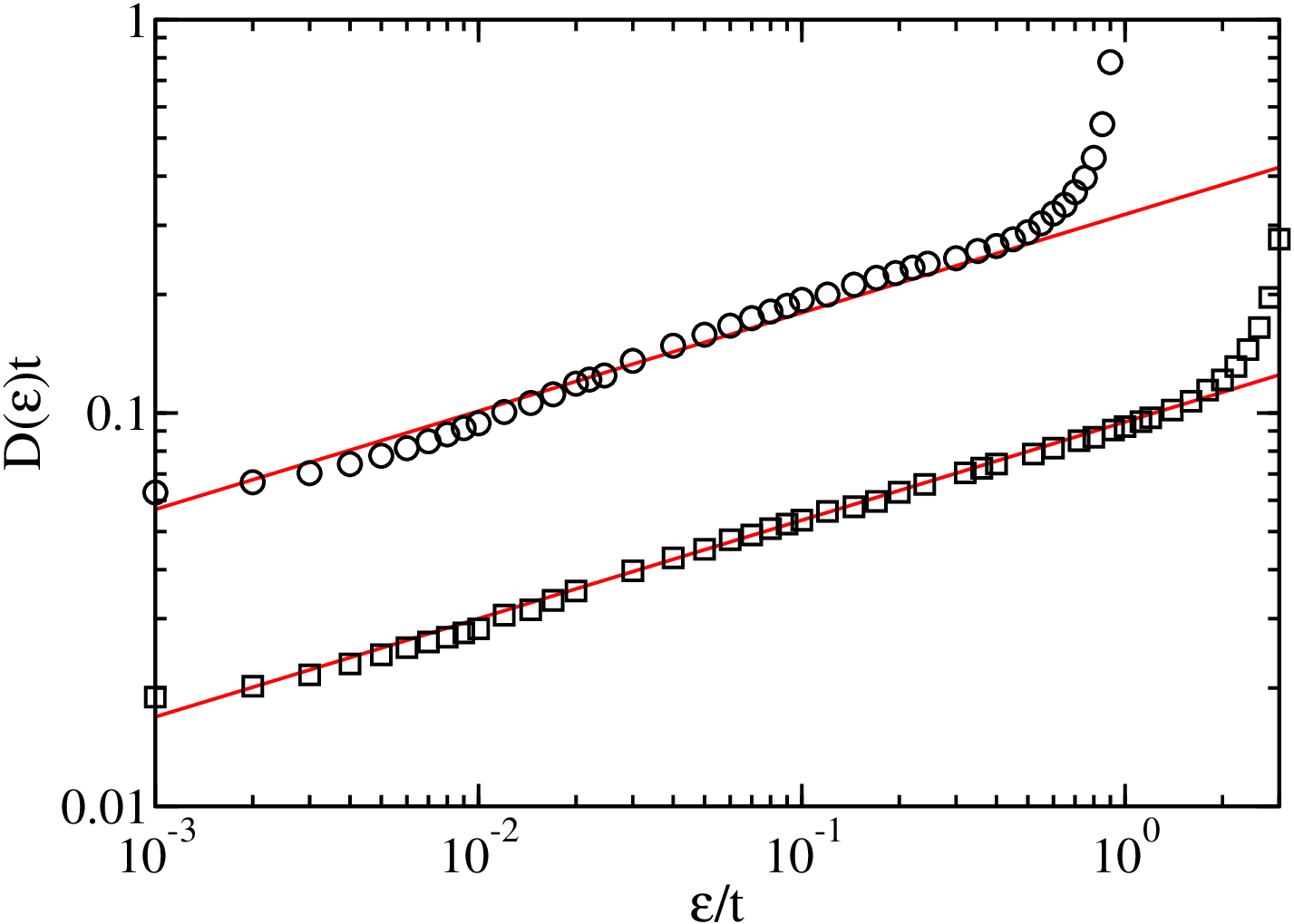}
\caption{\label{fig:logbulkdos} Bulk density of states for $V=-\sqrt{2}t$ (\opencircle)
and $V=2t$ (\opensquare) 
calculated with DDMRG in a 1600-site chain ($\eta=0.01t$).
The solid lines show the power law~\eref{eq:LLdos} with the exact exponent 
$\alpha=\frac{1}{4}$ for both cases.}
\end{figure}

Actually, we have often found an approximate power-law behaviour 
over a rather broad energy range of the order of $t^{*}$ when the exponent
$\alpha$ is not too large.
This is illustrated in \fref{fig:logbulkdos} which shows the bulk density of states 
for two values of $V$ on a double logarithmic scale.
In both cases the DDMRG data clearly deviate from the power law at high energy
$\epsilon \gtrsim t^{*}$.
This is not surprising as the curvature of the dispersion~\eref{eq:dispersion}
becomes relevant above this energy scale.
The DDMRG data can also deviate from the power law at low energy 
$\epsilon \lesssim \eta =0.01t$,
where the spectrum is dominated by the Lorentzian broadening~\eref{eq:convolution},
although this is not apparent in both examples in~\fref{fig:logbulkdos}.
For $V=2t$ the DDMRG spectrum agree particularly well with the power law.
For most values of $V$, however, there is only a rather rough agreement
as shown by the second example, $V=-\sqrt{2}t$. 

\begin{figure}[t]
\centering
\includegraphics[width=8cm]{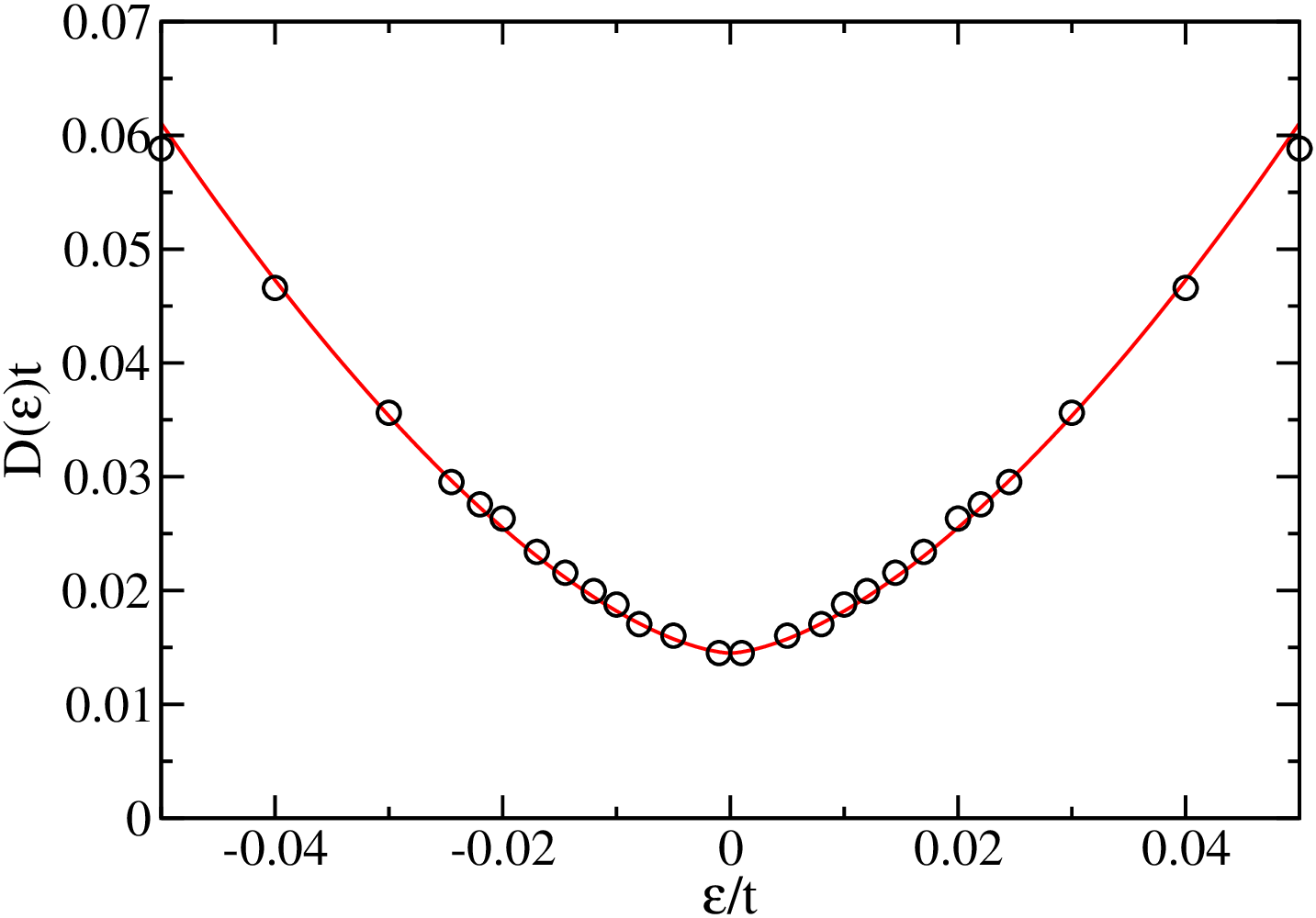}
\caption{\label{fig:expbulkdos2} Bulk density of states around the Fermi energy for
$V=-1.9t$ calculated with DDMRG in a 1600-site chain ($\eta=0.01t$).
The solid line is the power law~\eref{eq:LLdos} with the exact exponent 
$\alpha\approx 1.574$ but shifted vertically by 0.0145.}
\end{figure}

For large exponents $\alpha \gtrsim 1$  
discrepancies become apparently significant. This is illustrated
in~\fref{fig:expbulkdos2} which shows
the density of states for $V=-1.9t$ corresponding to $\alpha \approx 1.574$.
Clearly, a simple power law~\eref{eq:LLdos} can not describe the DDMRG data 
because they converge to a finite value for $\epsilon \rightarrow 0$.
This is due to the convolution~\eref{eq:convolution} which transfers
a spectral weight $D_{\rm DDMRG}(\epsilon_{\rm F}=0) \sim \eta \sim 1/N$ 
where the original spectrum has none.
While this finite-size effect occurs for all values of $V$, it becomes relevant
for large exponents $\alpha$ only.
Nevertheless, a simple shift is enough to offset the finite density of states
at the Fermi energy 
and to recover an excellent agreement with the low-energy power-law behaviour
as shown in~\fref{fig:expbulkdos2}.

We note that a similar effect is seen in the STS spectra of atomic 
chains~\cite{Blumenstein2011} and the purple bronze~\cite{Hager2005}. 
Clearly, the experimental densities of states shown in these publications 
are well fitted by power laws around the Fermi energy but they remain finite
at the Fermi energy. 
There, interchain couplings, finite temperature or disorder
may be responsible for this deviation from a pure
power-law behaviour.

\Fref{fig:logbulkdos} suggests that the bulk density of states decreases
approximately as a power law over a rather broad energy range.
Yet we have rarely succeeded in extracting accurate values for 
the exponent $\alpha$ from fits to such spectra.
The exponent $\alpha$ can be determined more accurately and with less computational effort
using a finite-size scaling analysis~\cite{Jeckelmann2002}.
For this purpose we calculate the density of states $D(\epsilon)$
for several system sizes $N$ 
at an energy $\epsilon$ which scales as $1/N$. 
We know that $D(\epsilon~\sim 1/N)$ scales as 
$N^{-\alpha}$ in the thermodynamic limit and thus we
can determine $\alpha$ using a power-law fit.
In this work we have used $\epsilon = \eta = 16t/N$
and up to $N=3200$ sites for this scaling analysis.
The exponents that we have obtained in the range $0.08 < \alpha < 1.6$
agree within 10\% with the predictions of the Luttinger liquid theory~\eref{eq:exponent1}
combined with the Bethe Ansatz solution~\eref{eq:Kparameter}.
Larger exponents than $\alpha \approx 1.6$
can be achieved in the half-filled spinless fermion chain
for $-1.9t > V > -2t$. We have not explore this regime because
the effective energy scale~\eref{eq:hopping} becomes rapidly very small
and thus numerical simulations become increasingly difficult.
A reliable determination of smaller exponents than $\alpha \approx 0.08$ 
would require
DDMRG computations for a larger range of system sizes than in this work 
(i.e., $N \gg 10^3$).
In principle, a similar approach could be used to determine 
whether the peaks found at $\epsilon = \pm 2t^{*}$ are 
band edge singularities like in the non-interacting system~\eref{eq:bulkdos0}
or smoother structures
but we have not carried out this analysis yet.

\section{Local density of states}

We now examine the LDOS $D(j,\epsilon)$ close to a chain end (hard wall boundary).
This configuration can also be realized experimentally, for instance in gold chains on Ge 
surfaces~\cite{Blumenstein2011}.
Previous theoretical works~\cite{Giamarchi,Schoenhammer2000,Meden2000,Eggert1996,Schneider2010}
have shown that in a Luttinger liquid 
the LDOS at chain edges differs significantly from the bulk density of states.
In particular, the low-energy behaviour is very different for 
spinless fermions with attractive ($V < 0$) or repulsive ($V > 0$) interactions. 
Additionally, the LDOS of the spinless fermion model next to a site impurity
has been investigated using the functional renormalization group (fRG) method
~\cite{Andergassen2004}. 
(In a Luttinger liquid the LDOS next to an impurity, however weak, is similar to the 
LDOS close to a chain end in the asymptotic low-energy limit~\cite{Giamarchi,Schoenhammer04}.)

\begin{figure}[b]
\centering
\includegraphics[width=8cm]{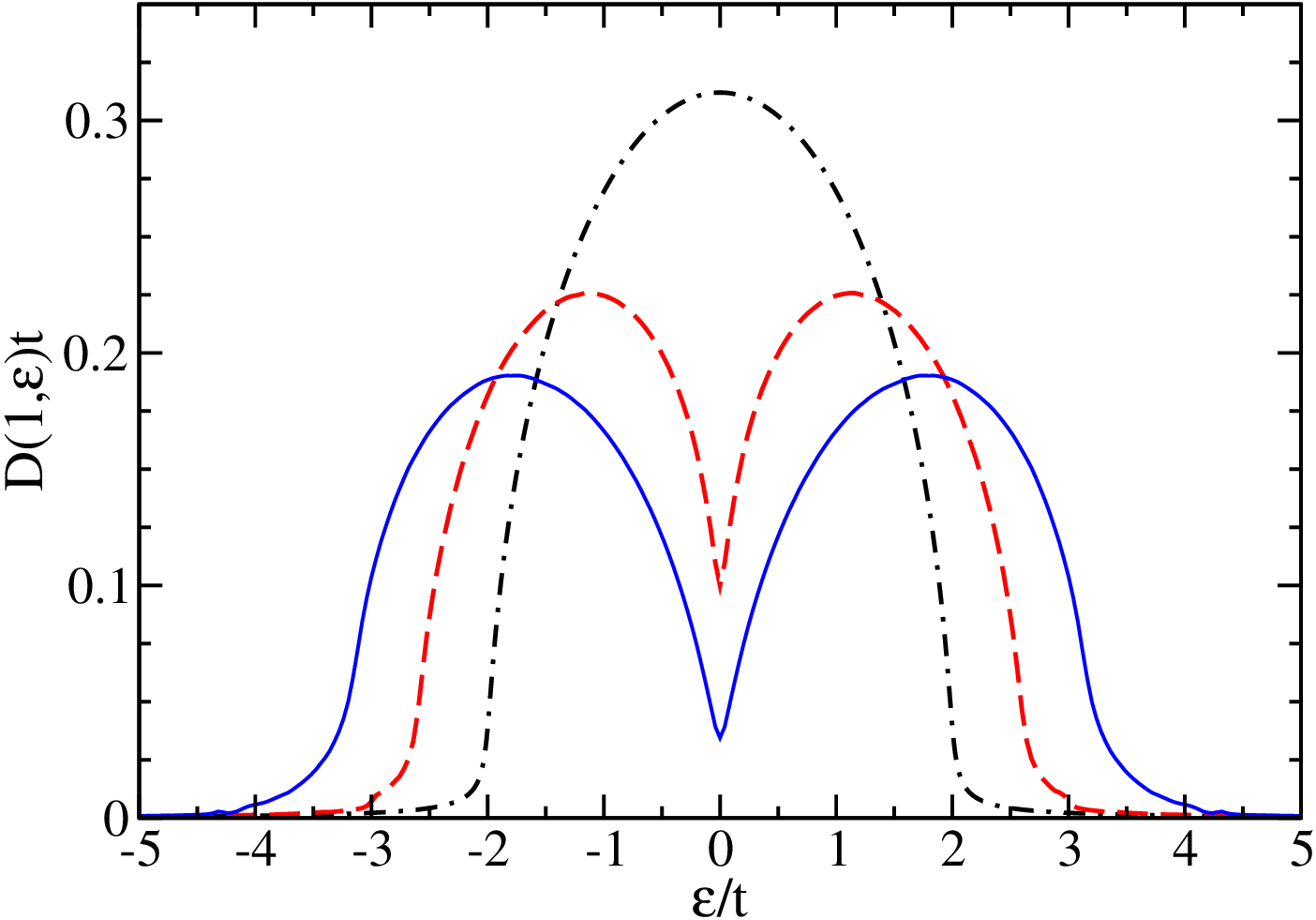}
\caption{\label{fig:edgedosrepulsive} Boundary LDOS $D(j=1,\epsilon)$
calculated with DDMRG on the first site of a 400-site chain ($\eta=0.04t$) 
for $V=0$ (\chain), $V=1t$ (\dashed), and $V=2t$ (\full).}
\end{figure}

We first explore the boundary LDOS, i.e $D(j=1,\epsilon)$ on the first lattice site.
In a semi-infinite non-interacting chain ($V=0$) this LDOS is
given by
\begin{equation}
D(j=1,\epsilon) = \frac{1}{2\pi t^2} \sqrt{4t^2-\epsilon^2} . 
\label{eq:boundarydos0}
\end{equation}
The DDMRG spectrum for $V=0$ is shown in \fref{fig:edgedosrepulsive}
with a broadening $\eta=0.04t$. On the scale of this figure
there is no visible difference between these DDMRG data
and the exact result~\eref{eq:boundarydos0} with the same broadening. 
This confirms that the scaling of the broadening~\eref{eq:eta}
is appropriate for the LDOS close to chain edges too and demonstrates again the accuracy
of the DDMRG method.

\Fref{fig:edgedosrepulsive} also shows the boundary LDOS calculated
with DDMRG for two repulsive interactions $V$. The semi-elliptic 
spectrum~\eref{eq:boundarydos0} of the non-interacting chain
seems to split into two smaller semi-elliptic bands which 
move apart as $V$ increases. A narrow minimum (but no gap up to $V=2t$)
appears at the Fermi energy $\epsilon=0$. The onsets
of the upper and lower bands correspond to the peak positions $\epsilon = \pm 2t^{*}$
in the bulk density of states discussed in the previous section
and thus to the edges of the elementary excitation bands~\eref{eq:dispersion}.
As in the bulk density of states we observe some spectral weight beyond $2t^{*}$
which can be explained by the convolution~\eref{eq:convolution} for $V=0$
but not for interacting fermions ($V \neq 0$).

\begin{figure}[t]
\centering
\includegraphics[width=8cm]{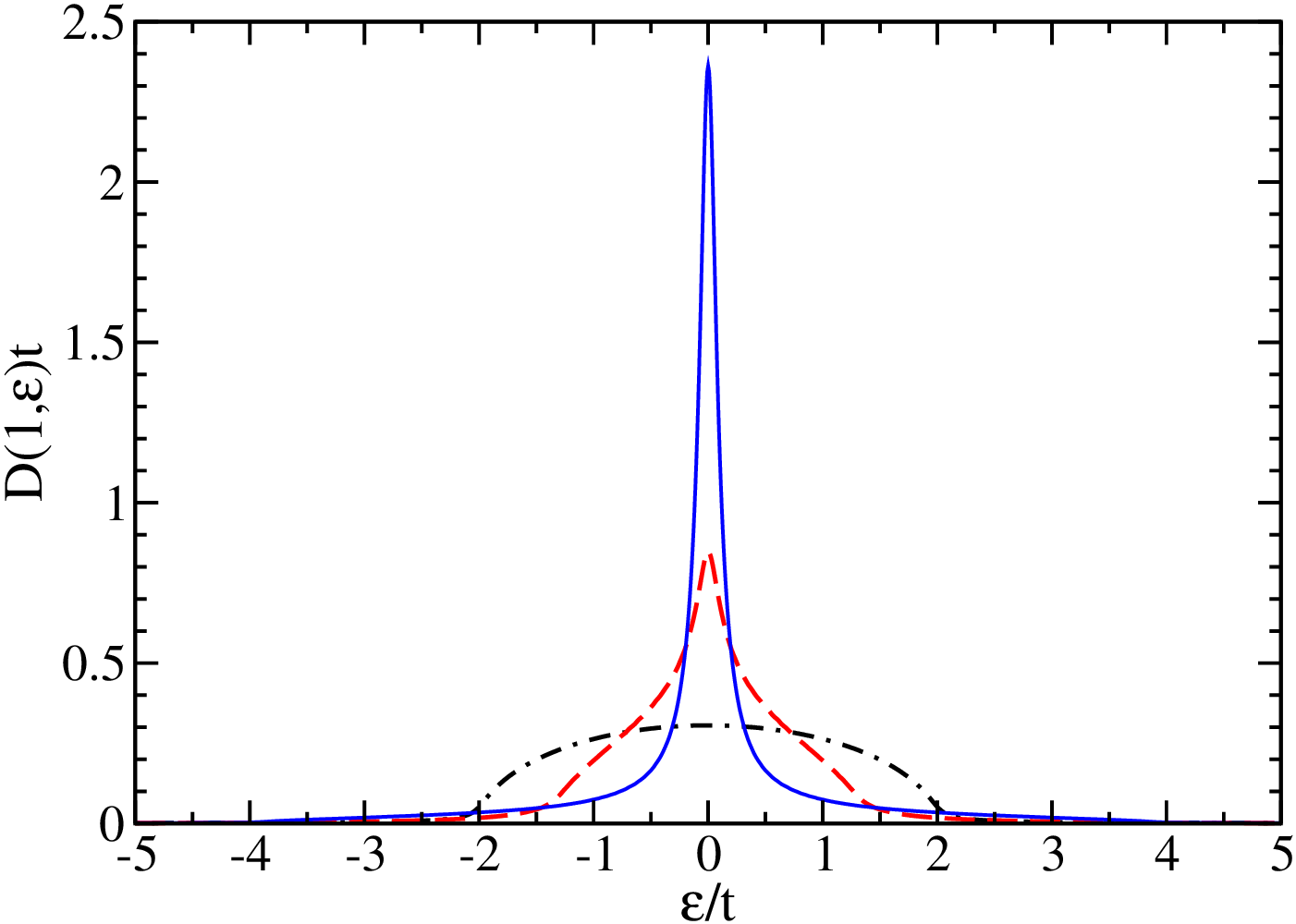}
\caption{\label{fig:edgedosattractive} Boundary LDOS $D(j=1,\epsilon)$
calculated with DDMRG on the first site of a 200-site chain ($\eta=0.08t$) 
for $V=0$ (\chain), $V=-1t$ (\dashed), and $V=-1.9t$ (\full).}
\end{figure}

In contrast for attractive interactions $V < 0$ the boundary
LDOS develops an increasingly sharp peak at the Fermi energy
as $V$ decreases. \Fref{fig:edgedosattractive} shows
the DDMRG spectra for two values of $V <0$ as well as $V=0$ for comparison.
For $V = -t$ we can still see that the spectrum has apparent onsets
at energies $\epsilon \approx \pm 2t^{*}$ corresponding
to the band edges of elementary excitations~\eref{eq:dispersion}.
For larger $V$ we can no longer distinguish the spectrum onsets
but most of the spectral weight
is still concentrated within the band of elementary 
excitations $-2t^{*} \leq \epsilon \leq 2t^{*}$.

The Luttinger liquid theory
predicts that the boundary density of states also follows a power law~\eref{eq:LLdos}
but the boundary exponent is 
\begin{equation}
\alpha_{\rm B}= \frac{1}{K} - 1 
\label{eq:exponent2}
\end{equation}
for a one-component Luttinger liquid~\cite{Giamarchi,Schoenhammer2000}.
Previous DMRG investigations
have confirmed
this expression for the boundary exponent in terms
of the Luttinger parameter~\eref{eq:Kparameter} 
obtained from the Bethe Ansatz solution~\cite{Schoenhammer2000,Meden2000}.

For repulsive interactions $V > 0$ the
Luttinger parameter is $K < 1$ according to~\eref{eq:Kparameter}
and thus $\alpha_{\rm B} > \alpha > 0$.
This increase of the exponent at a chain boundary has been observed
experimentally in gold chains on semiconducting Ge surfaces~\cite{Blumenstein2011}.
The power-law behaviour is not visible on the scale of \fref{fig:edgedosrepulsive}
and the narrow minima at the Fermi energy just hints at the presence of a pseudogap.
As for the bulk density of states we have examined the low-energy LDOS with
resolution of $\eta = 0.01t$ and obtained similar results. The boundary LDOS
follows an approximate power law over an energy range $\sim t^{*}$ for
small exponents $\alpha$. For $\alpha \approx 1$ ($\Leftrightarrow V \approx 2t$)
we have to offset some finite spectral weight at $\epsilon_{\rm F} = 0$
 to recover the correct low-energy behaviour.
Finally, using a finite-size scaling analysis of our DDMRG data
we obtain exponents $\alpha$ which agree within 10\% with the exact results
of the Luttinger liquid theory~\eref{eq:exponent2} combined with the Bethe
Ansatz solution~\eref{eq:Kparameter}. 

In contrast for attractive interactions $V < 0$ the
Luttinger parameter is $K > 1$ and thus $\alpha_{\rm B} < 0$.
Therefore, the Luttinger liquid theory predicts a power-law divergence
in the boundary LDOS at the Fermi energy. 
As the effective band width $\sim 4t^{*}$ and the exponent $\alpha_{\rm B}$ 
decrease rapidly with decreasing $V$,
this singularity should become stronger and sharper.
When $V$ approaches $-2t$, the peak width must vanish according to~\eref{eq:hopping}
while the exponent $\alpha_{\rm B}$ converges to -1 according to~\eref{eq:Kparameter} 
and~\eref{eq:exponent2}.
Thus the power-law singularity turns into a Dirac $\delta$-peak at the Fermi energy in that limit.
The predictions of the Luttinger liquid theory
agree with the boundary LDOS calculated with DDMRG which are shown
in~\fref{fig:edgedosattractive}. 
In the DDMRG spectra the Fermi-energy singularity has been smoothed into a peak 
by the broadening~\eref{eq:convolution}. 
Again we can perform a finite-size scaling analysis to confirm that
the peak is a singularity and to determine the precise values 
of the exponent $\alpha_{\rm B}$.
This analysis is much easier for power-law singularities
with negative exponents (i.e., divergences) than with positive ones
(pseudogaps). 
Thus we find exponents which agree very well (within 2\%)
with the exact values given by the 
Luttinger liquid exponent~\eref{eq:exponent2}
combined with the Bethe Ansatz solution~\eref{eq:Kparameter}.

We now turn to the crossover from the boundary LDOS to the bulk density of states.
First, it is helpful to discuss the LDOS close to the chain
edge in a semi-infinite non-interacting chain ($V=0$).
Numbering the sites $j=1,2, \dots$ from the chain edge
we obtain for small $j\ll N$  and $\vert \epsilon \vert < 2t$
\begin{equation}
D(j,\epsilon) = \frac{2}{\pi} \frac{1}{\sqrt{4t^2-\epsilon^2}}  \left [ 
1 - T^2_j \left ( \frac{\epsilon}{2t} \right ) \right ]
\label{eq:edgeldos0}
\end{equation}
where $T_j(x) = \cos(\arccos(x) j)$ are the Chebyshev polynomials.
As for the bulk density of states $D(j,\epsilon) = 0$ for $\vert \epsilon \vert > 2t$.
The boundary LDOS~\eref{eq:boundarydos0} is recovered for $j=1$.
The LDOS $D(j,\epsilon)$ oscillates between 0 and twice its bulk value~\eref{eq:bulkdos0}.
There are exactly $j$ peaks in $D(j,\epsilon)$ considered as a function of $\epsilon$.
In particular, there is a peak at the Fermi energy for odd $j$
but $D(j,\epsilon)$ vanishes as $\epsilon^2$ for even $j$.
At low energy these oscillations correspond to two modes $(-1)^j$
and $\cos(\epsilon j/t) = \cos[2\epsilon j a/(\hbar v_F)]$.
Field theoretical studies have shown that these same two oscillation 
modes also dominate the low-energy LDOS of Luttinger liquids
with the renormalized velocity $v$ substituted for $v_F$~\cite{Eggert1996}.

As~\eref{eq:edgeldos0} has been derived for $j \ll N$, it is not surprising that it does not 
converge toward the bulk density of states~\eref{eq:bulkdos0} for large $j$. 
Nevertheless, if we broaden each LDOS $D(j,\epsilon)$ with a Lorentzian~\eref{eq:convolution}
(or a Gaussian, \dots) of width $ \sim 4t/j$, we find that the result converges toward
the bulk density of states for $j\rightarrow \infty$. As in the middle of a $N$-site chain 
this broadening $\sim 4t/j$ is smaller than the one used in DDMRG calculations~\eref{eq:eta}, 
this property explains why the LDOS calculated in the middle of a chain with DDMRG agrees
so well with the (broadened) bulk density of states~\eref{eq:bulkdos0}.

\begin{figure}[t]
\centering
\includegraphics[width=4cm]{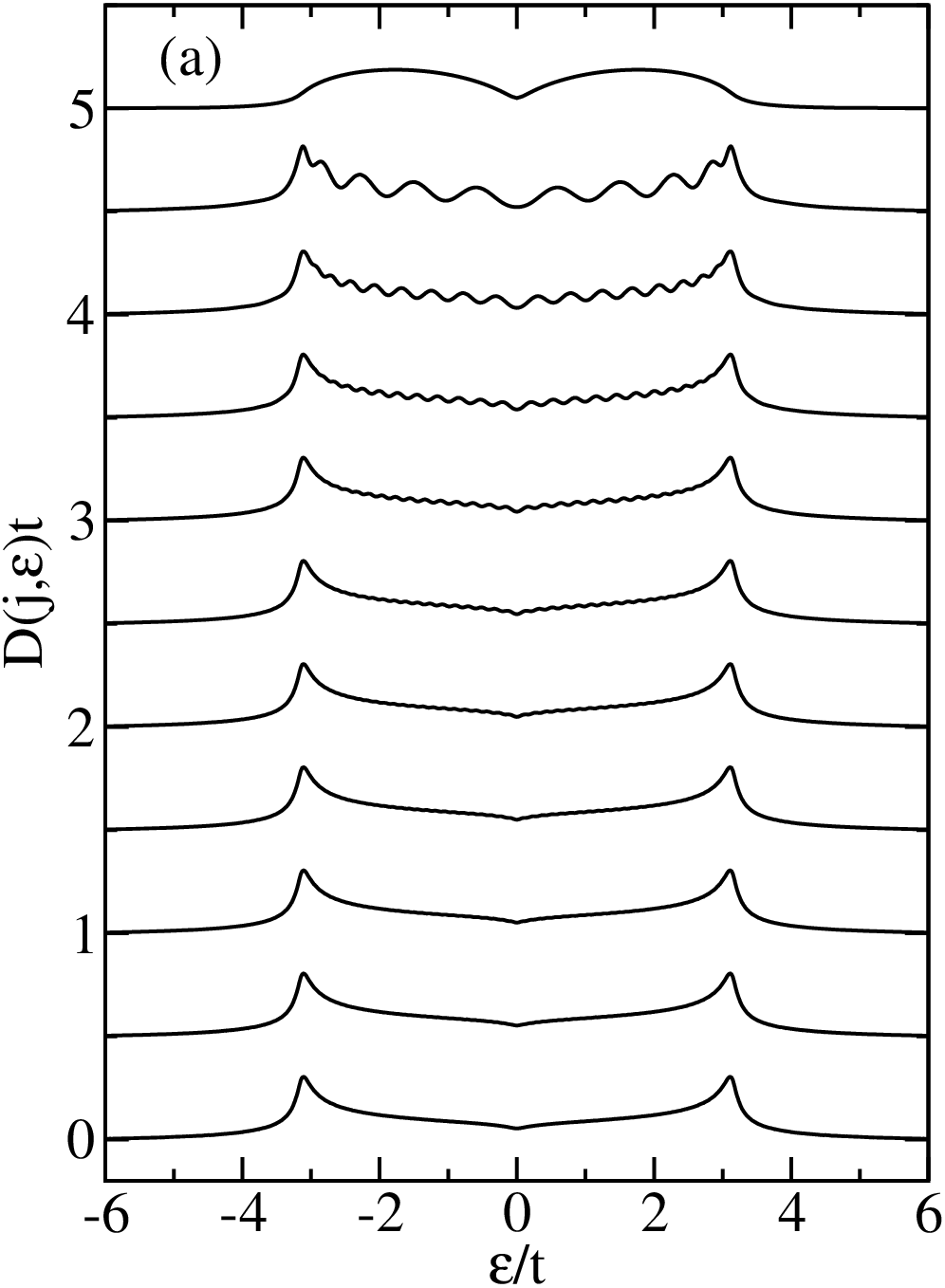}
\includegraphics[width=4cm]{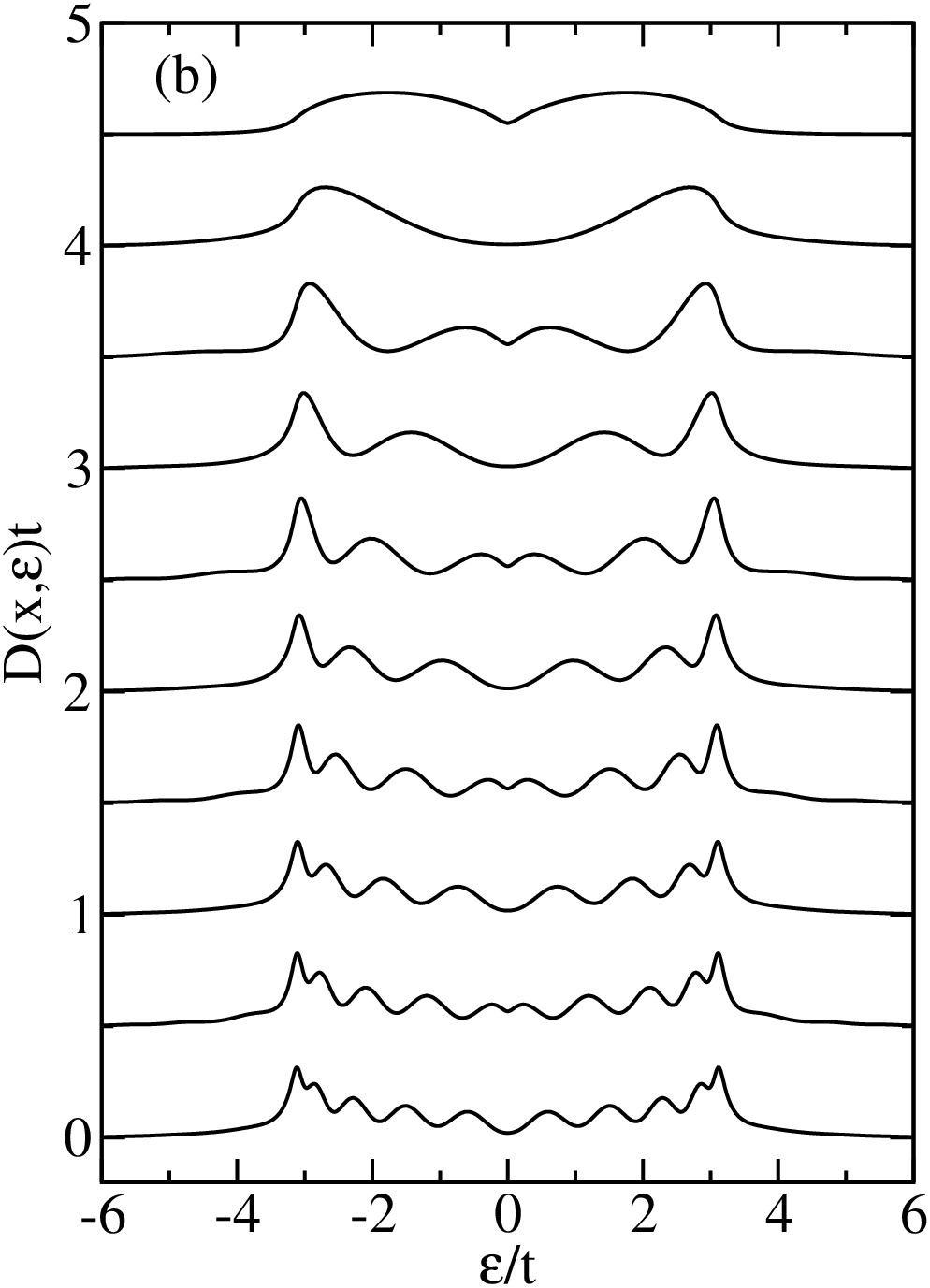}
\caption{\label{fig:localdosrepulsive} LDOS $D(j,\epsilon)$ 
calculated on a 200-site chain ($\eta=0.08t$) for $V=2t$.
(a) From one edge to the middle of the chain: $j=1,10,20,\dots,100$ from top to bottom.
(b) Close to the chain edge: $j=1,2,\dots,10$ from top to bottom.}
\end{figure}

\begin{figure}[t]
\centering
\includegraphics[width=4cm]{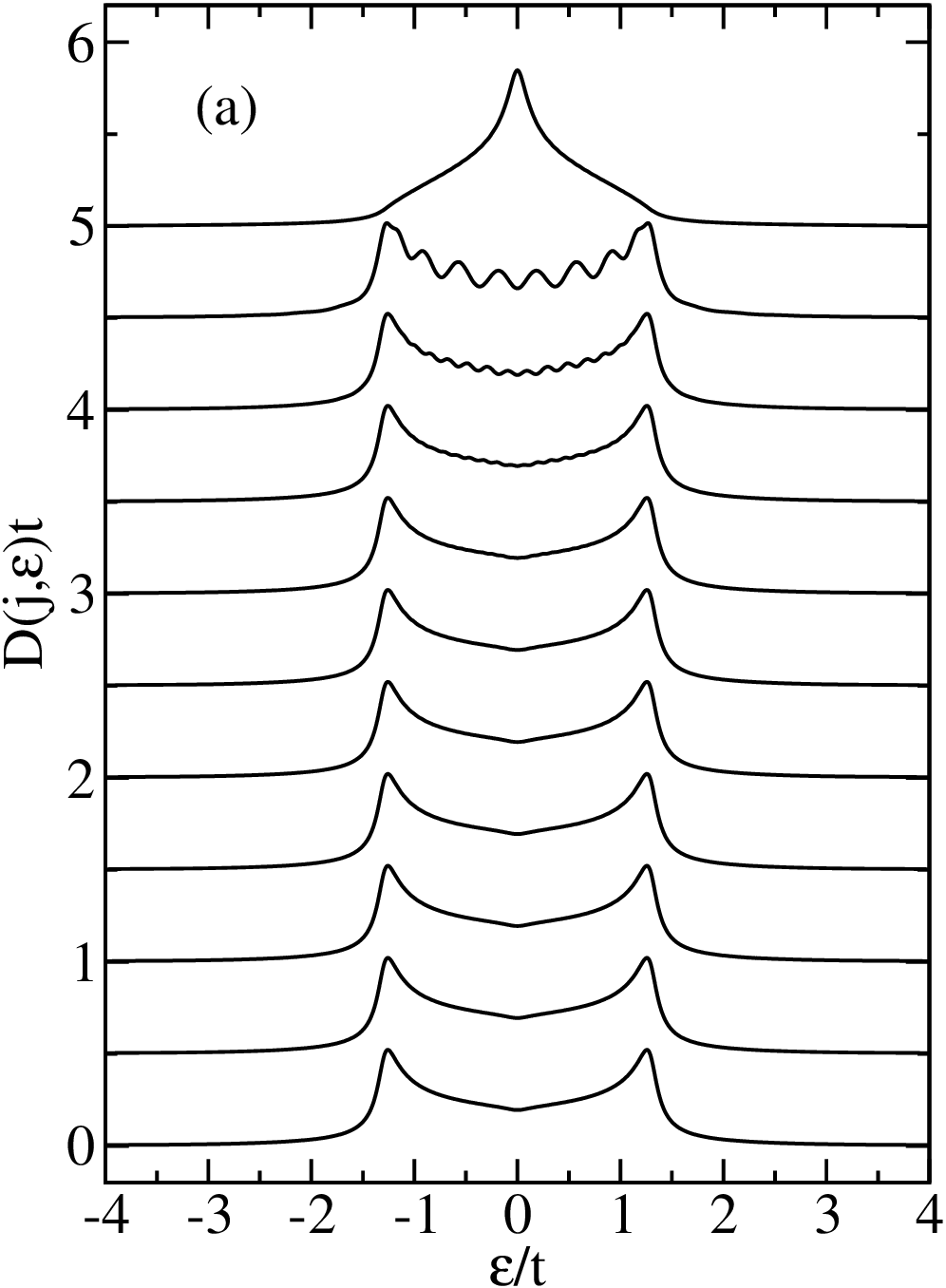}
\includegraphics[width=4cm]{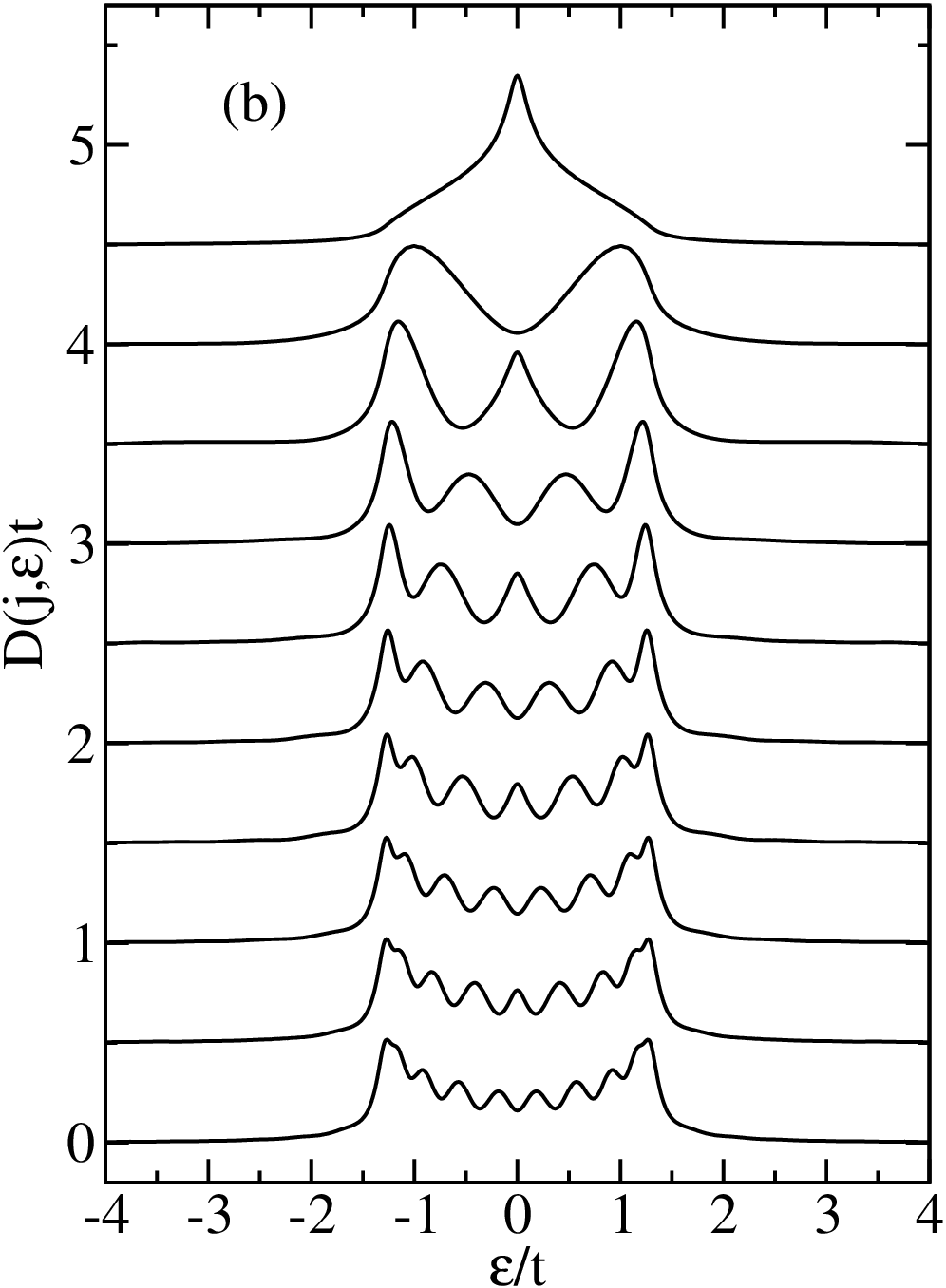}
\caption{\label{fig:localdosattractive} LDOS $D(j,\epsilon)$
calculated on a 200-site chain ($\eta=0.08t$) for $V=-t$.
(a) From one edge to the middle of the chain: $j=1,10,20,\dots,100$ from top to bottom.
(b) Close to the chain edge: $j=1,2,\dots,10$ from top to bottom.}
\end{figure}

The crossover from boundary to bulk density of states is shown in~\fref{fig:localdosrepulsive} 
for  a repulsive interaction $V=2t$
and in~\fref{fig:localdosattractive} for an attractive interaction $V=-1t$.
First, the right panels of both figures reveal that the characteristic shapes
of the boundary LDOS $D(j=1,\epsilon)$ in figures~\ref{fig:edgedosrepulsive}
and~\ref{fig:edgedosattractive} are purely local features.
The LDOS $D(j > 2,\epsilon)$ on other sites look completely different, even on the closest sites. 
Therefore, it is unlikely that the boundary exponent~\eref{eq:exponent2} 
can be determine precisely in STS experiments which average over several \AA~\cite{Blumenstein2011}.
Second, we note that the LDOS for $j>1$
are quite similar for attractive and repulsive interactions besides the
different
band width $\approx 4t^{*}$. 
The main reason for this likeness is that the spectra $D(j,\epsilon)$  
are clearly dominate by $j$ peaks up to (at least) $j =10$ for all interactions $V$. 
These oscillations are similar to those found in the 
LDOS for non-interacting spinless fermions~\eref{eq:edgeldos0}.
Finally, in the left panels of figures \ref{fig:localdosrepulsive} and~\ref{fig:localdosattractive} we
see that the amplitude of these oscillations decrease rapidly with increasing distance $j$ from the boundary.
Already for  $j \gtrsim 50 = N/4$ both LDOS do not appear to change on the scale of 
the broadening $\eta=0.08t$.
This confirms that DDMRG calculations of the LDOS in the middle of a chain yield  
a (broadened) bulk density of states.

The above discussion is based on our DDMRG data in chains with up to 200 sites 
and $\eta \geq 0.08t$.
As seen in the study of bulk and boundary density of states, 
this resolution is not good enough to analyze the 
asymptotic behaviour close to the Fermi energy.
Moreover, the multiple oscillations observed in the LDOS in the crossover regime $1 < j \ll N$  
imply that a power-law behaviour could only be observed in a significantly smaller energy range than
for the bulk and boundary density of states.
We estimate that carrying out such a study would require a 
computational effort which is one order of magnitude higher than for the
present work (which amounts to $2 \cdot 10^{4}$ CPU hours).

The crossover from boundary to bulk density of states in Luttinger liquids 
has already been
studied with field theory~\cite{Eggert1996,Schneider2010}.
We have not performed any quantitative comparison between DDMRG spectra
and the field-theoretical results or the fRG results~\cite{Andergassen2004}
because of the limited resolution of our data.
Nevertheless, there is a clear discrepancy between our results and one of the main 
predictions of bosonization.
In an extension of the Tomonaga-Luttinger model with hard wall boundaries  it has been shown that 
boundaries affect the LDOS $D(j,\epsilon)$ only below a crossover energy~\cite{Eggert1996}.
Above this energy bulk behaviour is observed.
This crossover energy is given by $v\hbar/(ja)=2t^{*}/j$.
However, in DDMRG spectra  (see figures~\ref{fig:localdosrepulsive} and~\ref{fig:localdosattractive}) 
boundary effects are clearly seen as rapid oscillations over the full band width, even for $j=20$.
Thus our results indicate that there is no crossover energy above which only 
bulk behaviour is observed in the spinless fermion model.
This is also demonstrated by the exact LDOS of noninteracting fermions~\eref{eq:edgeldos0} 
which can be regarded as a Luttinger liquid with $K=1$. 
A correct statement for the spinless fermion model is that boundary effects are only visible in 
the LDOS $D(j,\epsilon)$ at a resolution  $\lesssim 4t^{*}/j$.
If the LDOS is examined at a resolution larger than this value (for instance, if
it has been convolved with a Lorentzian or Gaussian distribution of width $\gtrsim 4t^{*}/j$),
then boundary effects are not visible.
Thus in the DDMRG spectra boundary effects disappear when the broadening~\eref{eq:eta}
exceeds a value $\approx 4t^{*}/j$.
 
\section{Conclusion}

We have investigated the local density of states (LDOS) $D(j,\epsilon)$ 
of the one-dimensional half-filled spinless fermion 
model in its Luttinger liquid phase.
Using the DDMRG method we have determined the LDOS over its full energy range
with a resolution of $\eta=0.08t$ or better.
In all cases most of the spectral weight is concentrated in the band 
$\vert \epsilon \vert  < 2t^{*}$ 
of elementary excitations from the Bethe Ansatz solution.   
We have found that the bulk density of states can be identified with the LDOS computed in the
middle of a chain if we use a large enough broadening $\eta$.
It is dominated by sharp peaks at the band edges $\epsilon = \pm 2t^{*}$
for all interactions $V$.
For repulsive interactions ($V > 0$) the boundary LDOS (on the first lattice site) 
consists in two symmetric joined bands which are reminiscent of Hubbard bands. 
For attractive interactions ($V < 0$) the boundary LDOS is dominated by
a peak at the Fermi energy, which becomes increasingly sharp as $V$ decreases
and turns into a Dirac $\delta$-peak for $V\rightarrow -2t$.

We have also examined the bulk and boundary density of states
close to the Fermi energy
with a resolution $\eta=0.01t$ on lattices with 1600 sites.
In the bulk density of states for $V\neq 0$ and in the boundary LDOS for $V>0$
we can observe the power-law pseudogap $D(j,\epsilon) \sim \vert \epsilon\vert^{\alpha}$
predicted by the Luttinger liquid theory over an energy range $\sim t^{*}$
although we have to compensate explicitly for finite-size effects when the
exponent $\alpha$ is large.  
In the boundary LDOS for $V<0$ we see the power-law divergence
predicted by the Luttinger liquid theory.
Using a finite-size scaling analysis of DDMRG data on lattices
with up to 3200 sites
we obtain exponents $\alpha$ which agree quantitatively with the exact results
given by the Luttinger liquid theory combined with the Bethe Ansatz solution.

Finally, we have explored the crossover from the boundary LDOS $D(j=1,\epsilon)$ 
to the bulk density of states $D(j=N/2,\epsilon)$ 
as one moves away from the chain edge. The characteristic
features of the boundary LDOS $D(j=1,\epsilon)$ in Luttinger liquids
appear to be a purely local effect as all $D(j>2,\epsilon)$ look completely different.
They are dominated by oscillations
which are similar to those observed in the non-interacting chain.
The crossover is smooth and boundary effects can be visible
far away from the chain edge ($j \gg 1$) at all energies $\epsilon$
but vanish beyond a point set by the DDMRG resolution,
i.e. when $4t^{*}/j \lesssim \eta$.

The LDOS of the spinless fermion model cannot be used directly
to interpret photoemission or STS experiments.
For this purpose we have to consider electronic models such as
the Hubbard model.
The present work has been motivated in part by the failure to observe a
power law in the density of states of the Hubbard model
using DDMRG with an energy resolution $\eta=0.1t$.
For spinless fermion models the DDMRG method allows us to study  larger 
system sizes $N$ and thus to reach a better resolution $\sim 1/N$
than for electronic models.
Our work confirms that the power-law features of Luttinger liquid LDOS
can be studied with DDMRG. However, the resolution required for the Hubbard
model is certainly much smaller than $0.1t$.
In the spinless fermion model we have to use a resolution of about $10^{-1} t^{*}$
for $\alpha \approx 1$ and $10^{-2} t^{*}$ for $\alpha \approx \frac{1}{4}$
to observe a power law.
In the Hubbard model $\alpha < 1/8$ and the energy scale of
the power-law behaviour is much lower than the bare band width $4t$.
This scale is exponentially small $\sim \exp(-t/U)$ for small $U/t$~\cite{Meden2000} and 
it is probably set by the effective exchange coupling
$J \sim t^2/U$ for large $U/t$.  
Thus a resolution even smaller than $0.01t$ is required
which implies system sizes well in excess of $10^3$ sites.

The recent experimental observation of Luttinger liquid LDOS
in atomic wires certainly calls for a renewed effort in the study of the LDOS
in Hubbard-type models for interacting electrons in one dimension.
Fortunately, larger exponents $\alpha$ are achieved 
in extended Hubbard models with a non-local Coulomb interaction~\cite{Ejima2005,Ejima2006},
which are anyway more realistic than the original Hubbard model.
We think that in these electronic models the LDOS and the power-law pseudogap 
could be investigated with a sufficient accuracy and a reasonable 
computational effort using the DDMRG method.

\ack
I thank Volker Meden for helpful comments.
The GotoBLAS library developed by Kazushige Goto was used to perform the DDMRG calculations.
Some of these calculations were carried out on the RRZN cluster system
of the Leibniz Universit\"at Hannover.


\section*{References}
\begin{thebibliography}{99}
\bibitem{Baeriswyl} Baeriswyl D and Degiorgi (eds) 2004
{\it Strong Interactions in Low Dimensions} (Dordrecht: Kluwer Academic Publishers) 
\bibitem{Giamarchi} Giamarchi T 2003 {\it Quantum Physics in One Dimension} (Oxford: Clarendon Press)
\bibitem{Giuliani} Giuliani G F and Vignale G 2005 {\it Quantum Theory of the Electron Liquid}
(Cambridge: Cambridge University Press)
\bibitem{Schoenhammer02} Sch\"{o}nhammer K 2002 \JPCM \textbf{14} 12783
\bibitem{Schoenhammer04} Sch\"{o}nhammer K 2004 Luttinger liquids: the basic concepts
{\it Strong Interactions in Low Dimensions} ed. D Baeriswyl and L Degiorgi 
(Dordrecht: Kluwer Academic Publishers) 
\bibitem{Vescoli2000} Vescoli V \etal 2000 \textit{Eur. Phys. J.} B \textbf{13} 503
\bibitem{Sing2003} Sing M  \etal 2003 \PR B \textbf{68} 125111
\bibitem{Ishii2003} Ishii H \etal 2003 \textit{Nature} \textbf{426} 540
\bibitem{Hager2005} Hager J  \etal 2005 \PRL \textbf{95} 186402 
\bibitem{Wang2006} Wang F  \etal 2006 \PR B \textbf{74} 113107 
\bibitem{Blumenstein2011} Blumenstein C \etal 2011 \textit{Nature Physics} \textbf{7} 776
\bibitem{Imambekov2009} Imambekov A and Glazman L I 2009 \textit{Science} \textbf{323} 228
\bibitem{Essler05} Essler F H L, Frahm H, G\"{o}hmann F, Kl\"{u}mper A and Korepin V E 
2005 {\it The One-Dimensional Hubbard Model} (Cambridge: Cambridge University Press)
\bibitem{Kohno2010} Kohno M, Arikawa M, Sato J and Sakai K 2010 \textit{J. Phys. Soc. Jap.}
\textbf{79} 043707
\bibitem{Fehske} Fehske H, Schneider R and Wei\ss e A (eds) 2008 {\it Computational Many-Particle Physics}
(Berlin: Springer)
\bibitem{Schollwoeck2005} Schollw\"ock U 2005 \RMP \textbf{77} 259
\bibitem{Schoenhammer2000} Sh\"onhammer K, Meden V, Metzner W, Schollw\"ock U and Gunnarsson O
2000 \PR B \textbf{61} 4393
\bibitem{Meden2000} Meden V, Metzner W, Schollw\"ock U, Schneider O, Stauber T and Sch\"onhammer K
2000 \textit{Eur. Phys. J.} B \textbf{16} 631
\bibitem{Schneider2008} Schneider I, Struck A, Bortz M and Eggert S 2008 \PRL \textbf{101}
206401
\bibitem{Jeckelmann2002} Jeckelmann E 2002  \PR B \textbf{66} 045114
\bibitem{Jeckelmann2008b} Jeckelmann E and Benthien H 2008 Dynamical Density-Matrix 
Renormalization Group {\it Computational Many-Particle Physics} ed. 
H Fehske, R Schneider and A Wei\ss e (Berlin: Springer)
\bibitem{Jeckelmann2008} Jeckelmann E 2008 \textit{Progress of Theoretical Physics 
Supplement} \textbf{176} 143 
\bibitem{Caux2011} Caux J-S, Konno H, Sorrell M and Weston R 2011 \PRL \textbf{106} 217203
\bibitem{Eggert1996} Eggert S, Johannesson H and Mattsson A 1996 \PRL \textbf{76} 1505
\bibitem{Schneider2010} Schneider I and Eggert S 2010 \PRL \textbf{104} 036402
\bibitem{Andergassen2004} Andergassen S, Enss T, Meden V, Metzner W, Schollw\"ock U and
Sch\"onhammer K 2004 \PR B \textbf{70} 075102
\bibitem{Ejima2005} Ejima S, Gebhard F and Nishimoto S 2005 \textit{Europhysics Letters} \textbf{70}
492
\bibitem{Ejima2006} Ejima S, Gebhard F and Nishimoto S 2006 \PR B \textbf{74} 245110 
\end{thebibliography}
\end{document}